\documentclass[aps,preprint,showpacs,preprintnumbers,eqsecnum,amsmath,amssymb]{revtex4}

\usepackage{graphics,epsfig}
\usepackage{graphicx}
\usepackage{dcolumn}
\usepackage{bm}

\begin{document}

\title{Deformation of contour and Hawking temperature}
\author{Chikun Ding \footnote{Email: dingchikun@163.com} and Jiliang Jing
}\thanks{Corresponding author, Electronic address:
jljing@hunnu.edu.cn}
 \affiliation{ Institute of Physics and
Department of Physics,
Hunan Normal University, Changsha, Hunan 410081, P. R. China \\
and
\\ Key Laboratory of Low Dimensional Quantum Structures and
Quantum Control of Ministry of Education, Hunan Normal University ,
Changsha, Hunan 410081, P.R. China}

 \baselineskip=0.65 cm

\vspace*{0.2cm}
\begin{abstract}
\vspace*{0.2cm}

It was found that,  in an isotropic coordinate system, the tunneling
approach brings a factor of $\frac{1}{2}$ for the Hawking
temperature of a Schwarzschild black hole. In this paper, we address
this kind of problem by studying the relation between the Hawking
temperature and the deformation of integral contour for the scalar
and Dirac particles tunneling. We find that correct Hawking
temperature can be obtained exactly as long as the integral contour
deformed corresponding to the radial coordinate transform if the
transformation is a non-regular or zero function at the event
horizon.

\vspace*{0.2cm}

Keywords:  black hole, Hawking temperature, integral contour.

\end{abstract}

 \vspace*{1.5cm}
 \pacs{04.70.Dy, 04.62.+v}

\maketitle

\section{introduction}

A semi-classical Hamilton-Jacobi method \cite{man3}-\cite{wil4} for
controlling Hawking radiation as a tunneling effect has been
developed recently. In this method a semiclassical propagator
$K(\vec{x}_2,t_2;\vec{x}_1,t_1)$ in a spacetime is described by $
N\exp\big[\frac{i} {\hbar}(I(\vec{x}_2,t_2;\vec{x}_1,t_1) +C\big]$
in which the action $I(\vec{x}_2,t_2;\vec{x}_1,t_1)$ acquires a
singularity at the event horizon. This singularity can be
regularized by specifying a suitable complex contour \cite{man3}.
After integrating around the pole, we find that the action
$I(\vec{x}_2,t_2;\vec{x}_1,t_1)$ is complex. Thus, we know that the
probabilities are $ \Gamma[\text{emission}]\propto
e^{-2\text{Im}[I_++C]}$ and $
 \Gamma[\text{absorption}]\propto
e^{-2\text{Im}[I_-+C]}=1,$ and the ratio is
\begin{eqnarray}\label{gamma2}
\Gamma[\text{emission}]=e^{-2[\text{Im}I_+-\text{Im}I_-]}
\Gamma[\text{absorption}],\end{eqnarray}where $I_\pm$ are the square
roots of the relativistic Hamilton-Jacobi equation corresponding to
outgoing and ingoing particles. In a system with a temperature
$T_H$, the absorption and the emission probabilities are related by
$\label{gamma00}\Gamma[\text{emission}]=e^{-
E/T_H}\Gamma[\text{absorption}]$. Then, from the relation
\begin{eqnarray}\label{gamma0}
 e^{-E/T_H}=e^{-2[\text{Im}I_+-\text{Im}I_-]},\end{eqnarray}
we can obtain the Hawking temperature.

It is well known that the Hawking temperature is an attribution of
the black hole and is independent of coordinates. This can be seen
from its definition: $T_H=\frac{\kappa}{2\pi}$ \cite{beken}, where
$\kappa$ is the surface gravity of the black hole. However, to
calculation the Hawking temperature by tunneling approach, we need
to regularize the singularity by specifying a suitable complex
contour to bypass the pole. For the Schwarzschild black hole in the
standard coordinate representation, we should take the contour to be
an infinitesimal semicircle below the pole $r=r_H$ for outgoing
particles from inside of the horizon to outside; similarly, the
contour is above the pole for the ingoing particles from outside to
inside. But, if we use another coordinate representations, we find
that the calculation of the Hawking temperature is related to the
choice of the integral contour and improper contour would give
incorrect result. For example, if a semi-circular contour is still
employed in the isotropic coordinate system, the temperature
calculated by the Hamilton-Jacobi method is one-half of the standard
result (the so-called ``factor of $\frac{1}{2}$ problem", see
Appendix \ref{app2}); and if we use a semi-circular contour in a
general coordinate (\ref{arbitrary1}), we can prove that the
temperature would be $(\alpha+1)$ times of the standard result (see
Appendix \ref{app3}).

The `` factor of $\frac{1}{2}$ problem" of the Schwarzschild black
hole in the isotropic coordinates is studied by Aknmedov {\it et al}
\cite{akh1,akh2} by deforming the contour, i.e. using a
quarter-circular contour instead of the semi-circular contour. How
to extend it to a general case? In this manuscript we will study the
problem in a general coordinate \cite{ding,ding1} for a Kerr-Newman
black hole via the scalar and Dirac particles tunneling.

This paper is organized as follows. In Sec.II, the different
coordinate representations for the Kerr-Newman black hole are
presented.  In Sec. III,  the Hawking temperature of the Kerr-Newman
black hole from scalar particles tunneling in a general coordinate
is studied. In Sec. IV, the Hawking temperature of the Kerr-Newman
black hole from Dirac particles tunneling is studied. The last
section is devoted to a summary.

\section{Coordinate representations for a Kerr-Newman black hole}

The no-hair theorem postulates that all black hole solutions of the
Einstein-Maxwell equations of gravitation and electromagnetism in
general relativity can be completely characterized by only three
externally observable classical parameters: the mass, the electric
charge, and the angular momentum. The final state of a collapsing
star is described by the Kerr-Newman black hole. In the
Boyer-Lindquist coordinates, its line element reads
\begin{eqnarray}\label{kn}
&&ds^2=-\left(1-\frac{2Mr-Q^2}{\rho^2}\right)dt_s^2-\frac{2(2Mr-Q^2)a\sin^2\theta}
{\rho^2}dt_sd\varphi_s+\frac{\rho^2}{\triangle}dr^2\nonumber\\
&&\quad\quad\quad+\rho^2d\theta^2+\left(r^2+a^2+\frac{(2Mr-Q^2)a^2\sin^2\theta}{\rho^2}\right)\sin^2\theta
d\varphi_s^2,
   \end{eqnarray}with
   \begin{eqnarray}\nonumber
   &&\rho^2=r^2+a^2\cos^2\theta,~~~~ \triangle=r^2-2Mr+a^2+Q^2=(r-r_+)(r-r_-)\nonumber\\
   &&r_+=M+\sqrt{M^2-a^2-Q^2},~~~~r_-=M-\sqrt{M^2-a^2-Q^2},\nonumber
\end{eqnarray}
where $M$, $Q$ and $a$ are the mass, electric charge and angular
momentum of the black hole, and $r_-$ and $r_+$ are the inner and
outer horizons. The spacetime has a timelike Killing vector
$\tilde{\xi}^\mu_{(t)}=(1,0,0,0)$, and a spacelike Killing vector
$\tilde{\xi}^\mu_{(\varphi)}=(0,0,0,1)$.

We note that the Painlev\'{e}-type \cite{jiang}, advanced
Eddington-Finkelstein \cite{matt} and Boyer-Lindquist  coordinate
representations can be casted into an united form which is given by
a general coordinate transform
 \begin{eqnarray}\label{arbitrary}
u=\int dr F(r),~~~v=\eta t_s+\eta\int(r^2+a^2)G(r) dr ,~~~
\varphi=\delta\varphi_s+\delta a\int  G(r)dr,
\end{eqnarray}
where  ($t_s,~r,~ \theta, ~\varphi_s$) are the Boyer-Lindquist
coordinates; $v$, $u$ and $\varphi$ represent the time, radial and
angular coordinates respectively, $\theta$ remains the same; $\eta$
and $\delta$ are arbitrary nonzero constants which re-scale the time
and angle; and $G$ and $F$ are arbitrary functions of $r$ only. The
line element (\ref{kn}) in the new coordinate system becomes
\begin{eqnarray}\label{arbitrary1}
&&ds^2=-\frac{1}{\eta^2}\left(1-\frac{2Mr-Q^2}{\rho^2}\right)\bigg[
dv-\frac{\eta
G(r^2+a^2)}{F}du\bigg]^2\nonumber\\&&~~~~-\frac{2(2Mr-Q^2)a\sin^2\theta}
{\eta\delta\rho^2}\bigg[dv-\frac{\eta
G(r^2+a^2)}{F}du\bigg]\bigg[d\varphi-\frac{\delta a G}{F}du\bigg]\nonumber\\
&& ~~~~+\frac{\rho^2}{\triangle
F^2}du^2+\rho^2d\theta^2+\left(r^2+a^2+\frac{(2Mr-Q^2)a^2\sin^2\theta}{\rho^2}\right)
\frac{\sin^2\theta}{\delta^2}\bigg[d\varphi-\frac{\delta a
G}{F}du\bigg]^2.
   \end{eqnarray}
The timelike and spacelike Killing vectors of the spacetime are
\begin{eqnarray}\xi^\mu_{(t)}=\frac{\partial
x^\mu}{\partial\tilde{x}^\nu}\tilde{\xi}^\nu_{(t)}=(\eta,0,0,0),\quad
\xi^\mu_{(\varphi)}=\frac{\partial
x^\mu}{\partial\tilde{x}^\nu}\tilde{\xi}^\nu_{(\varphi)}=(0,0,0,\delta).\end{eqnarray}

 {\it{Painlev\'{e}-type coordinate representation}}:
In the transformation (\ref{arbitrary}), if we take $\eta=\delta=1$,
$G(r)=\frac{1}{\triangle}\sqrt{\frac{2Mr-Q^2}{r^2+a^2}}$ and
$F(r)=1$, the line element (\ref{arbitrary1}) becomes the
Painlev\'{e}-type coordinate representation  \cite{jiang}, which has
no coordinate singularity at $\triangle(r)=0$.

 {\it{Advanced Eddington-Finkelstein coordinate representation}}:
In the transformation (\ref{arbitrary}), if we let
$\eta=1,~\delta=-1$, $G(r)=\frac{1}{\triangle}$ and $F(r)=1$, the
line element (\ref{arbitrary1}) becomes the advanced
Eddington-Finkelstein representation, which has no coordinate
singularity just as in the Painlev\'{e}-type coordinates
\cite{matt}.

{\it{Boyer-Lindquist  coordinate representation}}: In the
transformation (\ref{arbitrary}), if we let $\eta=\delta=F(r)=1$,
$G(r)=0$, the line element (\ref{arbitrary1}) becomes the
Boyer-Lindquist  coordinate representation (\ref{kn}).

\vspace*{0.4cm}

 \section{Temperature of Kerr-Newman black hole from
 scalar tunneling in the general coordinate system}\label{General}

Now we study the scalar tunneling in the general  coordinates
(\ref{arbitrary1}). Applying the WKB approximation
\begin{eqnarray}\label{ans}
\phi(v,u,\theta,\varphi)
=\exp\Big[\frac{i}{\hbar}I(v,u,\theta,\varphi)+I_1(v,u,\theta,\varphi)
+\mathcal{O}(\hbar)\Big]
   \end{eqnarray}
to the charged Klein-Gordon equation\begin{eqnarray}\label{kg}
\frac{1}{\sqrt{-g}}(\partial_{\bar{\mu}}-\frac{iq}{\hbar}A_{\bar{\mu}})\big[\sqrt{-g}g^{\bar{\mu}\bar{\nu}}
(\partial_{\bar{\nu}}-\frac{iq}{\hbar}A_{\bar{\nu}})\phi\big]
-\frac{\mu^2}{\hbar^2}\phi=0,
   \end{eqnarray}
 then, to leading order in $\hbar$, we obtain the
 relativistic Hamilton-Jacobi equation
   \begin{eqnarray}\label{hj}
g^{\bar{\mu}\bar{\nu}}(\partial_{\bar{\mu}} I\partial_{\bar{\nu}}
I+q^2A_{\bar{\mu}}A_{\bar{\nu}}-2qA_{\bar{\mu}}\partial_{\bar{\nu}}
I)+\mu^2=0,
   \end{eqnarray}where $\mu$ is the mass of tunneling particles.
From the symmetries of the metric (\ref{arbitrary1}), we know that
there exists a solution of the form (see Appendix \ref{app})
 \begin{eqnarray}\label{action1}
I=-\frac{1}{\eta}Ev+W(u)+\frac{1}{\delta}m\varphi+ J(\theta)+C.
\end{eqnarray}
Substituting the metric (\ref{arbitrary1}) and Eq. (\ref{action1})
into the Hamilton-Jacobi equation (\ref{hj}), we obtain
\begin{eqnarray}\label{separation3}
  &&\triangle^2 \left[F W'(u)-(r^2+a^2)G
  \Big(E-\frac{qQr}{r^2+a^2}-\frac{ma}{r^2+a^2}\Big)\right]^2
 \nonumber\\&& -\Big(r^2+a^2\Big)^2\left[
E-\frac{qQr}{r^2+a^2}-\frac{
ma}{r^2+a^2}\right]^2+\triangle\lambda=0,
\end{eqnarray}
with \begin{eqnarray} \lambda=\mu^2\rho^2+J'^2(\theta)
+\bigg(aE\sin\theta-\frac{m}{\sin\theta}\bigg)^2,
 \end{eqnarray}where $W'(u)=\frac{dW(u)}{du}$, and
$J'(\theta)=\frac{dJ(\theta)}{d\theta}$.  Then, $W'(u)$ can be
expressed as
\begin{eqnarray}\label{w(u)}
   &&W'_\pm(u)=\frac{G}{F}(r^2+a^2)
   \Big(E-\frac{qQr} {r^2+a^2}-\frac{ma}{r^2+a^2}\Big)
   \nonumber\\&&
\quad\quad\quad\pm\frac{1}
   {F\triangle}\sqrt{\big(r^2+a^2\big)^2
\Big(E-\frac{qQr}{r^2+a^2}-\frac{ma}{r^2+a^2}\Big)^2
-\triangle\lambda}.\end{eqnarray}One solution of  Eq. (\ref{w(u)})
corresponds to the scalar particles moving away from the black hole
(i.e. ``+" outgoing), and the other solution corresponds to
particles moving toward the black hole (i.e. ``-" incoming). Without
loss of generality, the function $G$ can be expressed as
$G(r(u))=\frac{A(r(u))} {\triangle(r(u))}+B(r(u))$, where $A(r(u))$
and $ B(r(u))$ are regular functions. Thus,  we have
\begin{eqnarray}\label{wu}
   &&\text{Im}W_\pm(u)=\text{Im}\int du \bigg[\bigg(\frac{B}{F}+\frac{A}{F\triangle}\bigg)
   (r^2+a^2)\Big(E-\frac{qQr}{r^2+a^2}-\frac{ma}{r^2+a^2}\Big)\nonumber\\ &&~~~~
\quad\quad\quad\pm\frac{1}
   {F\triangle}\sqrt{\big(r^2+a^2\big)^2
\Big(E-\frac{qQr}{r^2+a^2}-\frac{ma}{r^2+a^2}\Big)^2
-\triangle\lambda}~\bigg]~.\end{eqnarray} Imaginary part of the
action can only come from the pole at the horizon. We will work out
the integral in two cases: A) $F$ is a regular and non-zero function
at the horizon, and B) $F$ is a singular or zero function at the
horizon.

   \subsection{$F$ is a
   regular and non-zero function at the horizon}\label{regularF}

If $F$ is a regular and non-zero function at the horizon, using the
law of residue we obtain
   \begin{eqnarray}\text{Im}W_\pm(u)
   &=&\Big[A(r_+)
  \pm1\Big]\frac{r_+^2+a^2}
   {2(r_+-M)}\left( E-
   m\Omega_+-qV_+\right)\pi,\end{eqnarray}where $V_+=\frac{Qr_+}{r^{2}_++a^2}$
   is the electromagnetic potential, and $\Omega_+=\frac{a}{r^{2}_++a^2}$
   is the angular velocity.  Then, Eqs.
   (\ref{gamma2}) and (\ref{gamma0}) show us that
 the total probability is
\begin{eqnarray}\label{gamma4}
\Gamma &=&\exp\left[-2\pi\frac{r_+^2+a^2}
   {(r_+-M)}\left( E-
   m\Omega_+-qV_+\right)\right],
   \end{eqnarray}
and the Hawking temperature is
\begin{eqnarray}\label{HT}
T_H=\frac{r_+-M}{2\pi (r_+^2+a^2)}, \end{eqnarray} which is the same
as previous work \cite{hawking2,man,man3,jiang}.

\subsection{$F$ is a singular or zero function at the horizon}

 If $F$ is a non-regular or zero function at the horizon, without
loss of generality, we set $F=\triangle^\alpha X(r)$, where $\alpha$
is a non-zero constant and $X(r)$ is a regular and non-zero
function. Thus, Eq. (\ref{wu}) becomes\begin{eqnarray}\label{alpha0}
   &&\text{Im}W_\pm(u)=\text{Im}\int du \bigg[\bigg(\frac{B}{\triangle^\alpha X}
   +\frac{A}{\triangle^{\alpha+1}X}\bigg)
   (r^2+a^2)\Big(E-\frac{qQr}{r^2+a^2}-\frac{ma}{r^2+a^2}\Big)\nonumber\\ &&~~~~
\quad\quad\quad\pm\frac{1}
   {\triangle^{\alpha+1}X}\sqrt{\big(r^2+a^2\big)^2
\Big(E-\frac{qQr}{r^2+a^2}-\frac{ma}{r^2+a^2}\Big)^2
-\triangle\lambda}~\bigg]~.\end{eqnarray}From which we know
 \begin{eqnarray}\label{alpha1}
&&\text{Im}[W_+(u)-W_-(u)]\nonumber\\ &&=2\text{Im}\int du \frac{1}
   {\triangle^{\alpha+1}X}\sqrt{\big(r^2+a^2\big)^2
\Big(E-\frac{qQr}{r^2+a^2}-\frac{ma}{r^2+a^2}\Big)^2
-\triangle\lambda}~.\end{eqnarray} We now study two cases: 1)
$\alpha\neq-1$ and 2) $\alpha=-1$.

 \subsubsection{ $\alpha\neq-1$}

\begin{figure}[ht]
\begin{center}
\includegraphics[]{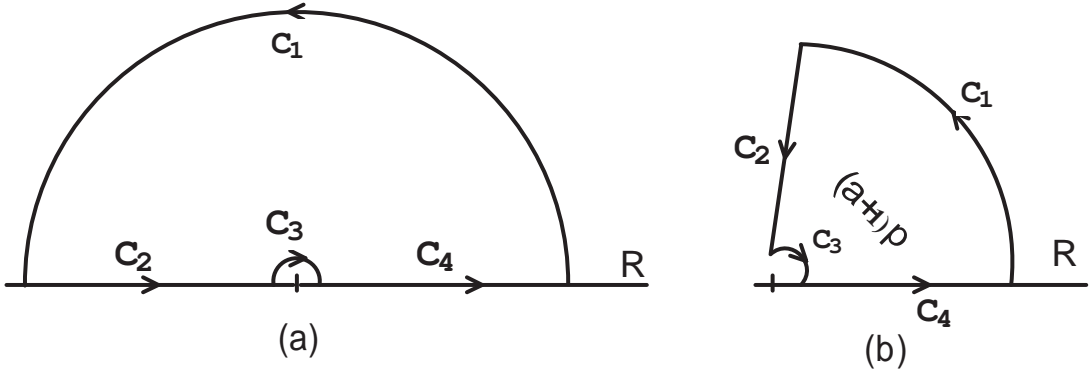}
\caption{The figure (a) is the semicircle integrate contour for
outgoing particles in the $r$ complex plane; and the figure (b) is
the deformation contour in the $u$ complex plane when
$\alpha\neq-1$. For $\alpha=-1$, its contour is still a semicircle.
The tick mark on the real axis denotes the position of the black
hole event horizon.}\label{figd}
\end{center}
\end{figure}

The Laurent expansion for the factor $\frac{1}{\triangle^
{\alpha+1}(r(u))}$ is
 \begin{eqnarray}
&&\frac{1}{\triangle^{\alpha+1}(r(u))}=
\frac{X(r(u_+))}{2(\alpha+1)(r_+-M)}\frac{1}{u-u_+}
+\sum^\infty_{n=0}a_n(u-u_+)^n.\end{eqnarray} Then Eq.
(\ref{alpha1}) can be written as \begin{eqnarray}\label{alpha}
&&\text{Im}[W_+(u)-W_-(u)]=2\text{Im}\int du\bigg[\frac{1}{\alpha+1}
\cdot\frac{1}{2(r_+-M)(u-u_+)}+\frac{1}{X}\sum^\infty_{n=0}a_n(u-u_+)^n\bigg] \nonumber\\
&&~~~~ \quad\quad\quad\cdot\sqrt{\big(r^2+a^2\big)^2 \Big(E-\frac{q
Q r}{r^2+a^2}-\frac{ma}{r^2+a^2}\Big)^2
-\triangle\lambda}~.\end{eqnarray} Now, we need to choose a contour
to bypass the pole $u=u_+$.  We note that, in the Boyer-Lindquist
coordinate, the contour can be constructed by taking $r=r_++\epsilon
e^{i\theta}$, ($\epsilon$ is a positive small real quantity,
$\theta\in[0,\pi]$ for outgoing particle, $\theta\in[\pi,2\pi]$ for
ingoing particle). Thus, in the general coordinate
(\ref{arbitrary1}), by substituting $r=r_++\epsilon e^{i\theta}$
into $u=\int \triangle^\alpha X(r)dr =\int [(r-r_+)(r-r_-)]^\alpha
X(r)dr$, we have
\begin{eqnarray}\label{contour}
u&=&\int [\epsilon e^{i\theta}(r_+-r_-+\epsilon e^{i\theta})]
^\alpha X(r_++\epsilon e^{i\theta})d\epsilon
e^{i\theta}\nonumber \\
&=&u_++f(u_+) \epsilon ^{\alpha+1}e^{i(\alpha+1)\theta},
 \end{eqnarray}
where $f(u_+)=\frac{(r_+-r_-)^\alpha X(r_+)}{\alpha+1} $.
 Eq. (\ref{contour}) indicates  that the
 contour is different from semi-circle now. The integral contours
 for outgoing particles
 corresponding to $r$ and $u$ complex plane are showed in figure
 (\ref{figd}).
Using Eqs. (\ref{alpha}), (\ref{contour}) and residue theorem, we
have
\begin{eqnarray}
&& \text{Im}[W_+(u)-W_-(u)]\nonumber\\&&=-2\text{Im}
 \lim_{\epsilon\rightarrow0}\int
 _\pi^{0}
id\theta\bigg[\frac{1}{2(r_+-M)} +\frac{f(u_+)(\epsilon
e^{i\theta})^{\alpha+1}(\alpha+1)}{X}
\sum^\infty_{n=0}a_nf^n(u_+)(\epsilon
e^{i\theta})^{(\alpha+1)n}\bigg] \nonumber\\
&& ~~~~~~~~~~~~\cdot\sqrt{\big[(r_++\epsilon
e^{i\theta})^2+a^2\big]^2 \Big[E-\frac{qQ(r_++\epsilon
e^{i\theta})+ma}{(r_++\epsilon e^{i\theta})^2+a^2}\Big]^2
-\triangle\lambda}~\nonumber\\
&&=\pi\frac{r_+^2+a^2}
   {(r_+-M)}\left( E-
   m\Omega_+-qV_+\right),\end{eqnarray}
which gives the Hawking temperature (\ref{HT}).

\subsubsection{$\alpha=-1$}

It is the tortoise-like coordinate transformation if $\alpha=-1$
\begin{eqnarray}u=\int X(r)\triangle^{-1}dr.
 \end{eqnarray}
By using $r=r_++\epsilon e^{i\theta}$, we know
\begin{eqnarray}\label{contour2}u=u_++i\theta g(u_+),\end{eqnarray}
where $g(u_+)=\frac{X(r_+)}{r_+-r_-} $.  Substituting it into Eq.
(\ref{alpha1}), we obtain \begin{eqnarray}
 \text{Im}[W_+(u)-W_-(u)]&=&-2\text{Im}\lim_{\epsilon\rightarrow0}\int
 _\pi^{0}
\frac{id\theta g(u_+)}{(r_++\epsilon
e^{i\theta})^2+a^2}\nonumber\\
&& \qquad\quad\cdot\sqrt{\big[(r_++\epsilon
e^{i\theta})^2+a^2\big]^2 \Big[E-\frac{qQ(r_++\epsilon
e^{i\theta})+ma}{(r_++\epsilon e^{i\theta})^2+a^2}\Big]^2
-\triangle\lambda}~\nonumber\\
&=&\pi\frac{r_+^2+a^2}
   {(r_+-M)}\left( E-
   m\Omega_+-qV_+\right),\end{eqnarray}
which also presents the Hawking temperature (\ref{HT}).

Above discussions show us that: i) the integral contour needs to be
deformed corresponding to the radial coordinate transformation if
this transformations are non-regular or zero at the event horizon;
ii) the Hawking temperature is invariant in the general coordinate
representation (\ref{arbitrary1}) for the scalar particle tunneling.

 \vspace*{0.4cm}

\section{Temperature of Kerr-Newman black hole from Dirac
particle tunneling }

In this section, we study the Dirac particle tunneling of the
Kerr-Newman black hole in the coordinates (\ref{arbitrary1}). The
Dirac equation is \cite{rmp}
\begin{eqnarray}\label{dirac}
&&\left[\gamma^\alpha
e^{\bar{\mu}}_\alpha(\partial_{\bar{\mu}}+\Gamma_{\bar{\mu}}-\frac{iq}{\hbar}A_{\bar{\mu}})+\frac{\mu}{\hbar}\right]
\psi=0,
 \end{eqnarray}
  with
  \begin{eqnarray}
&&\Gamma_{\bar{\mu}}=\frac{1}{8}[\gamma^a,\gamma^b]e^{\bar{\nu}}_ae_{b{\bar{\nu}};{\bar{\mu}}},
  \nonumber  \end{eqnarray}
where $\gamma^a$ is the Dirac matrix, and  $e^{\bar{\mu}}_a$ is the
   inverse tetrad defined by
     $\{e_a^{\bar{\mu}}\gamma^a,~~~e_b^{\bar{\nu}}\gamma^b\}=2g^{{\bar{\mu}}{\bar{\nu}}}
     \times1$. For the Kerr-Newman metric in the general coordinate system (\ref{arbitrary1}),
      the tetrad $e_a^{\bar{\mu}}$ can be
     taken as
\begin{eqnarray}\label{tetrad}
 &&e^v_a=\left(\begin{array}{cccc}
{\frac{\sqrt{\chi
-\triangle^2\eta^2(r^2+a^2)^2G^2}}{\rho\sqrt{\triangle}}}, & 0, &0,
& 0\end{array} \right), \nonumber\\ &&e^u_a=\left(
\begin{array}{cccc}-\frac{1}{\rho\sqrt{\triangle}} \frac{\triangle^2FG\eta(r^2+a^2)}
{\sqrt{\chi -\triangle^2\eta^2(r^2+a^2)^2G^2}},
&\frac{1}{\rho\sqrt{\triangle}}\frac{\triangle F\sqrt{\chi}}
   {\sqrt{\chi -\triangle^2\eta^2(r^2+a^2)^2G^2}}, &0,   &0\end{array}\right),
   \nonumber\\ &&e^\theta_a=\left(\begin{array}{cccc}0,  & 0,  & \frac{1}{\rho},    &0\end{array}\right),
   \nonumber\\ &&e^{\varphi}_a=\left(
\begin{array}{cccc}\frac{a\eta\delta}{\rho\sqrt{\triangle}} \frac{(2Mr-Q^2)
-\triangle^2 (r^2+a^2)G^2} {\sqrt{\chi
-\triangle^2\eta^2(r^2+a^2)^2G^2}}, &\frac{a\delta
G\sqrt{\triangle}}{\rho} \frac{\chi-(2 Mr-Q^2) \eta^2(r^2+a^2)}
   {\sqrt{\chi(\chi -\triangle^2\eta^2(r^2+a^2)^2G^2)}},  & 0, &
{\frac{\eta\delta\rho}{\sin\theta\sqrt{\chi}} }\end{array}\right), \nonumber\\
&&\chi=\eta^2\Big[\big(r^2+a^2\big)^2-\triangle
a^2\sin^2\theta\Big].
\end{eqnarray}Without loss of generality, we can choose the following ansatz
for spin up and spin down Dirac particles according to
\cite{weinberg},
   \begin{eqnarray}\label{psi}
  &&\psi_\uparrow=\bigg(\begin{array}{ccc}A(v,u,\theta,\varphi)\xi_\uparrow\nonumber\\
   B(v,u,\theta,\varphi)\xi_\uparrow\end{array}\bigg)
   \exp\big(\frac{i}{\hbar}I_\uparrow(v,u,\theta,\varphi)\big)
   =\left(\begin{array}{ccc}A(v,u,\theta,\varphi)\nonumber\\ 0\nonumber\\
   B(v,u,\theta,\varphi)\nonumber\\0\end{array}\right)
   \exp\big(\frac{i}{\hbar}I_\uparrow(v,u,\theta,\varphi)\big),\nonumber\\
   &&\psi_\downarrow=\bigg(\begin{array}{ccc}C(v,u,\theta,\varphi)\xi_\downarrow\nonumber\\
   D(v,u,\theta,\varphi))\xi_\downarrow\end{array}\bigg)
   \exp\big(\frac{i}{\hbar}I_\downarrow(v,u,\theta,\varphi)\big)
   =\left(\begin{array}{ccc}0\nonumber\\ C(v,u,\theta,\varphi)\nonumber\\
   0\\D(v,u,\theta,\varphi)\nonumber\end{array}\right)
   \exp\big(\frac{i}{\hbar}I_\downarrow(v,u,\theta,\varphi)\big),\nonumber\\
   \end{eqnarray}
where ``$\uparrow$" and ``$\downarrow$" represent the spin up and
spin down cases, and $\xi_{\uparrow}$ and $\xi_{\downarrow}$ are the
eigenvectors of $\sigma^3$. Inserting Eqs. (\ref{tetrad}) and
(\ref{psi}) into Eq. (\ref{dirac}), and employing the ansatz
\begin{eqnarray}I_\uparrow=-\frac{1}{\eta}Ev +W(u)+\frac{1}{\delta}m\varphi
   +J(\theta)+C, \end{eqnarray} to
the lowest order in $\hbar$,  we obtain
\begin{eqnarray}\label{aa111}
  \left[-e^v_0\frac{1}{\eta}\big(E-\frac{qQr}{\rho^2}\big)
  +e^u_0W'(u)+e^\varphi_0\frac{1}{\delta}\big(m-\frac{qQr}
  {\rho^2}a\sin^2\theta\big)+\mu \right]A\quad&&
   \nonumber\\\quad+B\left[e^u_1W'(u)+e^\varphi_1
   \frac{1}{\delta}\big(m-\frac{qQr}{\rho^2}a\sin^2\theta\big)\right]
   =0&&,\\ \label{bb111}
 B\left[e^\theta_2J'(\theta)
+ie^\varphi_3\frac{1}{\delta}\big(m-\frac{qQr}{\rho^2}a\sin^2\theta\big)
   \right]=0&&, \\
  \label{cc111}
  -\left[-e^v_0\frac{1}{\eta}\big(E-\frac{qQr}{\rho^2}\big)
  +e^u_0W'(u)+e^\varphi_0\frac{1}{\delta}\big(m-\frac{qQr}{\rho^2}a\sin^2\theta\big)-\mu \right]B
   \quad&&\nonumber\\ \quad\quad-A\left[e^u_1W'(u)
   +e^\varphi_1\frac{1}{\delta}\big(m-\frac{qQr}{\rho^2}a\sin^2\theta\big)\right]
   =0&&,\\ \label{dd111}  -A\left[e^\theta_2J'(\theta) +ie^{\varphi}_3
  \frac{1}{\delta} \big(m-\frac{qQr}{\rho^2}a\sin^2\theta\big)
   \right]=0&&.
   \end{eqnarray}
 Eqs. (\ref{bb111}) and (\ref{dd111})
both yield $\big[e^\theta_2J'(\theta)
+ie^{\varphi}_3\frac{1}{\delta}(m-\frac{qQr}{\rho^2}a\sin^2\theta)\big]=
0$,
 regardless of $A$ or $B$. Then substituting tetrad elements
(\ref{tetrad}) into (\ref{aa111})--(\ref{dd111}), after tedious
calculating, we obtain
\begin{eqnarray}
 &&\triangle^2 \left[F W'(u)-G(r^2+a^2)
 \Big(E-\frac{qQr}{r^2+a^2}-\frac{ma}{r^2+a^2}\Big)\right]^2
 \nonumber\\&& -\Big(r^2+a^2\Big)^2\left[
E-\frac{qQr}{r^2+a^2}-\frac{
ma}{r^2+a^2}\right]^2\nonumber\\&&+\triangle\left[\mu^2\rho^2+J'^2(\theta)
+\bigg(aE\sin\theta-\frac{m}{\sin\theta}\bigg)^2\right]=0,
 \end{eqnarray}
which is the same as Eq. (\ref{separation3}). Therefore, it is easy
to find the Hawking temperature (\ref{HT}).

The spin-down calculation is similar to the spin-up case discussed
above, and the result is the same.

\section{summary}

We firstly cast three well-known coordinate representations, i.e.
the Painlev\'{e}-type, advanced Eddington-Finkelstein and
Boyer-Lindquist coordinate representations for the Kerr-Newman black
hole, into an united and general coordinate representation
(\ref{arbitrary1}). Then, based on this coordinate representation,
we study the relation between the Hawking temperature and the
deformation of integral contour for the scalar and Dirac particle
tunneling. We find that correct Hawking temperature can be obtained
exactly as long as the integral contour deformed corresponding to
the radial coordinate transform if the transformation is a
non-regular or zero function at the event horizon.

\begin{acknowledgments}This work was supported by the
National Natural Science Foundation of China under Grant No.
10875040, the National Basic Research of China under Grant No.
2010CB833004, the FANEDD under Grant No. 200317, the Hunan
Provincial Natural Science Foundation of China under Grant No.
08JJ0001, the construct program of the key discipline in hunan
province, the Hunan Provincial Innovation Foundation for
Postgraduate;  and Construct Program of the National Key Discipline.
\end{acknowledgments}
\appendix
\vspace*{0.4cm}

\section{Improper choice of contour
in isotropic coordinates for Schwarzschild black hole}\label{app2}

Improper choice of integral contour can led to incorrect
temperature. In this section, we review the following process
mentioned in \cite{ag,mn}. By taking an isotropic coordinate
transformation
\begin{eqnarray}\label{isotropic} t\rightarrow t,~~r\rightarrow
\rho,~~\ln \rho=\int \frac{dr}{r\sqrt{1-\frac{2M}{r}}},
   \end{eqnarray} the line element of the Schwarzschild black
hole becomes\begin{eqnarray}
ds^2=-\left(\frac{2\rho-M}{2\rho+M}\right)^2dt^2+\left(\frac{2\rho+M}{2\rho}\right)^4d\rho^2
+\frac{(2\rho+M)^4}{16\rho^2}d\Omega^2.
   \end{eqnarray}
The horizon now is $\rho_H=M/2.$ Substituting it and
$\phi=e^{i[-Et+W(\rho)+J(\theta,\varphi)]}$ into Hamilton-Jacobi
equation (\ref{hj}), we obtain\begin{eqnarray}\label{isocontour}
 \text{Im}W_{\pm}(\rho)&=&\pm\text{Im}\left[ \int
\frac{(2\rho+M)^3d\rho}{4\rho^2(2\rho-M)}\sqrt{E^2-(\frac{2\rho-M}
{2\rho+M})^2(m^2+g^{ij}J_iJ_j)}~\right].
   \end{eqnarray}
Because the imaginary part of above integration comes from the pole
$\rho=M/2$, we only consider the integral around the pole. If we
still set a semi-circular contour bypass the pole, as we do in the
Schwarzschild coordinates, i.e. setting $\rho=M/2+\epsilon
e^{i\theta}, \theta\in[0,\pi]$, then Eq. (\ref{isocontour}) becomes
\begin{eqnarray}
 \text{Im}W_{\pm}(\rho)&=&\mp\text{Im}\lim_{\epsilon\rightarrow0}\left[ \int_\pi^{0}
\frac{4(M+\epsilon e^{i\theta})^3id\theta}{(M+2\epsilon
e^{i\theta})^2}\sqrt{E^2-(\frac{\epsilon e^{i\theta}} {M+\epsilon
e^{i\theta}})^2(m^2+g^{ij}J_iJ_j)}~\right]\nonumber\\&=&\pm4\pi ME.
   \end{eqnarray}
By using Eqs. (\ref{gamma2}) and (\ref{gamma0}),  the probability is
\begin{eqnarray}
\Gamma=\frac{\Gamma[\text{emission}]}{\Gamma[\text{absorption}]}=
\exp\Big[-4\text{Im}W_+\Big] =\exp\Big[-16\pi
ME\Big]=\exp\Big[-\frac{E}{T}\Big],
   \end{eqnarray}
and incorrect temperature is
\begin{eqnarray}T=\frac{1}{16\pi M}=\frac{1}{2}T_H,\end{eqnarray}
where $T_H$ is the Hawking temperature of the Schwarzschild black
hole. This is the so-called ``factor of $\frac{1}{2}$ problem".

\section{improper choice of integral contour gives incorrect
temperature  in the general coordinates (\ref{arbitrary1}) }
 \label{app3}
 In Eq. (\ref{alpha}),
if we still set a semi-circular contour to bypass the pole $u=u_+$,
i.e. $u=u_++\epsilon e^{i\theta}$, we obtain
   \begin{eqnarray}\text{Im}[W_+-W_-]&=&\frac{\pi}{\alpha+1}
\left[\frac{r_+^2+a^2}
   {(r_+-M)}\left( E-
   m\Omega_+-qV_+\right)\right],\end{eqnarray}
   and  the temperature would be
 \begin{eqnarray} T=(\alpha+1)T_H.\end{eqnarray}

\section{definition of radiating particle energy and angular
momentum}\label{app}

It is well known that, there are conservational quantities as long
as the spacetime possesses some certain symmetries. In the
Borer-Lindquist coordinate system, the line element (\ref{kn})
obviously has temporal-translational invariance and
$\varphi_s$-translational one, so we can define the particle energy
as $E=-\partial_{t_s} I$, and the particle angular momentum as
$m=\partial_{\varphi_s}I$. Thus, the action can be written as
$I=-Et_s+W(r)+m\varphi_s+J(\theta)$, which is essentially related to
the time-like Killing vector $\tilde{\xi}^\mu_{(t)}=(1,0,0,0)$ and
the space-like Killing one $\tilde{\xi}^\mu_{(\varphi)}=(0,0,0,1)$.

As mentioned in ref. \cite{hans}, the scalar product between
time-like Killing vector and particle four-momentum
$p^\mu=mdx^\mu/d\lambda$ is a constant for the particle moving along
geodesic, i.e.
\begin{eqnarray}\label{pp3}\xi_\mu p^\mu=\text{constant}.\end{eqnarray}
Furthermore,  this scalar product is also a constant in different
coordinate systems. Hence, these quantities can be used to define
particle energy \cite{ding2} and angular momentum in different
coordinate systems, i.e.
\begin{eqnarray}\label{energy0}E=-\xi^\mu_{(t)} p_\mu,\quad
m=\xi^\mu_{(\varphi)} p_\mu.\end{eqnarray} In the general
coordinates (\ref{arbitrary1}),
    the energy and angular momentum of test particle are
   \begin{eqnarray}\label{energy1}&&E=-\xi^\mu_{(t)} p_\mu=-\xi^\mu_{(t)} \partial_\mu I
   =-\eta\partial_v
   I,\nonumber\\&&m=\xi^\mu_{(\varphi)} p_\mu=\xi^\mu_{(\varphi)} \partial_\mu I
   =\delta\partial_\varphi
   I.\end{eqnarray}
Thus, the expression of the action can be taken as
\begin{eqnarray}I=-\frac{1}{\eta}Ev+W(u)+J(\theta)
+\frac{1}{\delta}m\varphi.\end{eqnarray}

\end{document}